\documentclass[12pt]{article}
\usepackage{graphicx}
\def\hybrid{\topmargin 0pt      \oddsidemargin 0pt
        \headheight 0pt \headsep 0pt
        \voffset=-0.5cm
        \textwidth 6.25in       
        \textheight 9.5in       
        \marginparwidth 0.0in
        \parskip 5pt plus 1pt   \jot = 1.5ex}
\catcode`\@=11
\def\marginnote#1{}

\newcount\hour
\newcount\minute
\newtoks\amorpm
\hour=\time\divide\hour by60
\minute=\time{\multiply\hour by60 \global\advance\minute by-\hour}
\edef\standardtime{{\ifnum\hour<12 \global\amorpm={am}%
        \else\global\amorpm={pm}\advance\hour by-12 \fi
        \ifnum\hour=0 \hour=12 \fi
        \number\hour:\ifnum\minute<10 0\fi\number\minute\the\amorpm}}
\edef\militarytime{\number\hour:\ifnum\minute<10 0\fi\number\minute}

\def\draftlabel#1{{\@bsphack\if@filesw {\let\thepage\relax
   \xdef\@gtempa{\write\@auxout{\string
      \newlabel{#1}{{\@currentlabel}{\thepage}}}}}\@gtempa
   \if@nobreak \ifvmode\nobreak\fi\fi\fi\@esphack}
        \gdef\@eqnlabel{#1}}
\def\@eqnlabel{}
\def\@vacuum{}
\def\draftmarginnote#1{\marginpar{\raggedright\scriptsize\tt#1}}
\def\draftlabel#1{{\@bsphack\if@filesw {\let\thepage\relax
   \xdef\@gtempa{\write\@auxout{\string
      \newlabel{#1}{{\@currentlabel}{\thepage}}}}}\@gtempa
   \if@nobreak \ifvmode\nobreak\fi\fi\fi\@esphack}
        \gdef\@eqnlabel{#1}}
\def\@eqnlabel{}
\def\@vacuum{}
\def\draftmarginnote#1{\marginpar{\raggedright\scriptsize\tt#1}}

\def\draft{\oddsidemargin -.5truein
        \def\@oddfoot{\sl preliminary draft \hfil
        \rm\thepage\hfil\sl\today\quad\militarytime}
        \let\@evenfoot\@oddfoot \overfullrule 3pt
        \let\label=\draftlabel
        \let\marginnote=\draftmarginnote
   \def\@eqnnum{(\theequation)\rlap{\kern\marginparsep\tt\@eqnlabel}%
\global\let\@eqnlabel\@vacuum}  }

\def\numberbysection{\@addtoreset{equation}{section}
        \def\theequation{\thesection.\arabic{equation}}}

\def\underline#1{\relax\ifmmode\@@underline#1\else
        $\@@underline{\hbox{#1}}$\relax\fi}

\def\titlepage{\@restonecolfalse\if@twocolumn\@restonecoltrue\onecolumn
     \else \newpage \fi \thispagestyle{empty}\c@page\z@
        \def\thefootnote{\fnsymbol{footnote}} }

\def\endtitlepage{\if@restonecol\twocolumn \else  \fi
        \def\thefootnote{\arabic{footnote}}
        \setcounter{footnote}{0}}  
\relax

\numberbysection
\hybrid

\newfont{\Bbb}{msbm10 scaled 1\@ptsize00}
\newfont{\Bbbb}{msbm7 scaled 1\@ptsize00}
\newcommand{\CC}{\mbox{\Bbb C}}
\newcommand{\CCC}{\mbox{\Bbbb C}}
\newcommand{\DD}{\mbox{\Bbb D}}
\newcommand{\DDD}{\raise-1pt\hbox{$\mbox{\Bbbb D}$}}
\newcommand{\HH}{\mbox{\Bbb H}}
\newcommand{\HHH}{\mbox{\Bbbb H}}

\newcommand{\UUU}{\raise-1pt\hbox{$\mbox{\Bbbb U}$}}

\newcommand{\z}{\raise-1pt\hbox{$\mbox{\Bbbb Z}$}}

\def\beq{\begin{equation}}
\def\eeq{\end{equation}}
\def\p{\partial}

\def\DD{{\sf D}}

\def\BB{{\sf B}}

\begin{document}
\begin{titlepage}

\title{Growth of fat slits and dispersionless KP
hierarchy}

\author{A.~Zabrodin
\thanks{Institute of Biochemical Physics,
4 Kosygina st., 119334, Moscow, Russia and ITEP, 25
B.Cheremushkinskaya, 117218, Moscow, Russia}}

\date{November 2008}
\maketitle

\begin{abstract}
A ``fat slit" is a compact domain in the upper half plane
bounded by a
curve with endpoints on the real axis and
a segment of the real axis between them. We consider
conformal maps of the upper half plane to the exterior
of a fat slit parameterized by harmonic moments of the
latter and show that they obey an infinite set of Lax
equations for the dispersionless KP hierarchy. Deformation
of a fat slit under changing a particular harmonic moment
can be treated as a growth process similar to the Laplacian
growth of domains in the whole plane. This construction
extends the well known link between solutions to the
dispersionless KP hierarchy and conformal maps
of slit domains in the upper half plane and provides
a new, large family of solutions.

\end{abstract}

\vfill

\end{titlepage}

\section{Introduction}

Parametric families of
conformal maps in 2D are known to
be closely related to long wave limits of
nonlinear integrable PDE's and their infinite hierarchies.
This observation was first made
in \cite{GT} for mappings of slit domains and then
extended to mappings of domains bounded by
Jordan curves in \cite{WZ}.
In both cases conformal maps
from a standard reference
domain (such as upper half plane, or unit disk)
to a domain of a varying shape serve as
Lax functions of an integrable hierarchy whose
flows are identified with variations of the conformal maps
described by an infinite set of Lax equations.
The integrable structures arising in this way are dispersionless
Kadomtsev-Petviashvili (dKP) and dispersionless 2D Toda
(dToda) hierarchies and, more generally, the universal Whitham
hierarchy first introduced \cite{KriW,TakTak} in an absolutely different
context with the aim to describe slow modulations
of exact solutions to
soliton equations.

The most promising progress along these lines
was achieved in geometric
and physical interpretation of the dToda hierarchy. The key
fact, established in \cite{MWWZ} and further elaborated in
\cite{KMWWZ,Z1}, is that
variations of domains under the Toda flows
go exactly according to the Darcy law specific
for growth processes of Laplacian type and viscous hydrodynamics
in the Hele-Shaw cell with zero surface tension (see, e.g.,
\cite{list,book}).

Although the dKP hierarchy is simpler
than the dToda hierarchy, its role in the theory of conformal maps
and Laplacian growth is not well understood.
The connection with
conformal maps observed in \cite{GT} gives a
geometric interpretation to only
rather special (degenerate) solutions of the dKP hierarchy
which are reductions to
systems of hydrodinamic type with a finite number of
degrees of freedom.
As is shown in \cite{GT} (see also \cite{YG,MMM}),
they are related to conformal maps of the
upper half plane with slits emanating from the real axis.

The aim of this paper is to demonstrate that
the geometric interpretation of the dKP hierarchy is not
limited by
domains of such special kind. We show that the same dKP hierarchy
is able to cover a much broader class of
domains which can be
obtained from the upper half plane by removing
not just an infinitely thin slit but
a whole compact piece (of a non-zero area and arbitrary shape)
attached to the real axis, which we call
a ``fat slit" to stress the analogy (Fig. 1).
There is an important difference, however.
The dKP-evolution of
usual slits is actually finite dimensional because
the slits are to be regarded as
arcs of fixed curves \cite{GT,YG},
so only their endpoints can move. In contrast, the
dKP-evolution of fat slits (given by the Lax equations)
takes place in an infinite
dimensional variety corresponding to changing their shape
in an arbitrary way.

Let us recall the Lax formulation of the dKP hierarchy.
Starting from a Laurent series
\beq\label{int1}
z(p)=p+\sum_{k=1}^{\infty} a_k p^{-k}
\eeq
one introduces the dependence on an infinite
number of ``times" $T_1, T_2, T_3, \ldots$ via
Lax equations
\beq\label{int2}
\frac{\p z(p)}{\p T_k}=\{ B_k(p),\, z(p)\}:=
\frac{\p B_k}{\p p}\frac{\p z}{\p T_1}-
\frac{\p B_k}{\p T_1}\frac{\p z}{\p p}\,,
\eeq
where the generators of the flows $B_k$ are
polynomials in $p$ of the form $B_k(p)=(z^k(p))_{\geq 0}$
(polynomial parts of $z^k(p)$).
The dKP hierarchy is an infinite system of nonlinear
PDE's for
$a_i$'s resulting from comparing coefficients
in front of different powers of $p$ in the Lax equations
or in an equivalent system of equations of the
Zakharov-Shabat type
\beq\label{int3}
\frac{\p B_j(p)}{\p T_k}-
\frac{\p B_k(p)}{\p T_j}+
\{B_j(p), \, B_k (p)\}=0
\eeq
for all $j,k \geq 1$ (the Poisson bracket is defined in
(\ref{int2})). The coefficients $a_i$ and the times $T_i$
are assumed to be real numbers.

\begin{figure}[t]
   \centering
        \includegraphics[angle=-00,scale=0.45]{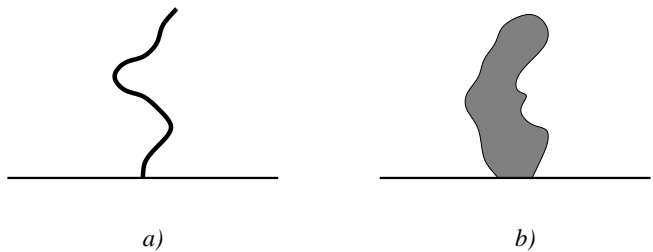}
        \caption{\it a) A slit; b) A ``fat slit".}
    \label{fig:fatslit}
\end{figure}

Assuming that $z(p)$ is a normalized
conformal map from the upper half plane onto the
exterior of a fat slit, we show that it obeys
equations (\ref{int2}), where $T_k$ are properly
defined harmonic moments of the fat slit (or rather
of its exterior). This fact follows from the
Hadamard formula for variations of the Green function
with Dirichlet boundary conditions.
In this sense the arguments are parallel
to \cite{KKMWZ,MWZ}. Evolution in $T=T_1$ with all other
times fixed has an interpretation as a version of
the Laplacian growth in the upper half plane with
fixed real axis.

Applying the approach developed in \cite{KKMWZ,MWZ,KMZ}
to the case of fat slits in the upper half plane,
we construct the dispersionless ``tau-function"
(which is actually a limit of properly
rescaled logarithm of a dispersionfull tau-function)
of the dKP hierarchy
as a functional
on the space of fat slits explicitly given by
\beq\label{int4}
\log \tau = -\, \frac{1}{\pi^2}
\int \!\!\!\!\!
\int\limits_{\mbox{{\tiny fat slit}}}^{}\!\! \log \left |
\frac{z-\zeta }{z-\bar \zeta}\right | \, d^2z d^2 \zeta \,.
\eeq
It has a clear electrostatic interpretation as
Coulomb energy of a fat slit filled with electric charge
of a uniform density in the presence of an infinite
grounded conductor placed along the real axis.
This functional regarded as a function of
harmonic moments obeys a dispersionless version of the
Hirota relation which serves as a master equation
generating the whole dKP hierarchy.

\section{Fat slit domains, their conformal maps and
Green's functions}

Consider a compact simply connected domain $\BB$ in the
upper half-plane $\HH$ bounded by a non-self-intersecting
analytic curve $\gamma$ in $\HH$ with endpoints
$x_{-}$, $x_{+}$ on the real axis and
a segment of the real axis between them. This
segment will be called
the {\it base} of $\BB$. Without loss of generality,
one can assume that the origin belongs to the base. For brevity, and
in order to emphasize an analogy with slit domains manifested
in the common integrable structure of their conformal maps, we call
such a domain {\it a fat slit}. Accordingly, the complement,
$\HH \setminus \BB$, will be referred to as a domain with
a fat slit,
or simply a fat slit domain (in our case, the fat slit half-plane).

\begin{figure}[t]
   \centering
        \includegraphics[angle=-00,scale=0.6]{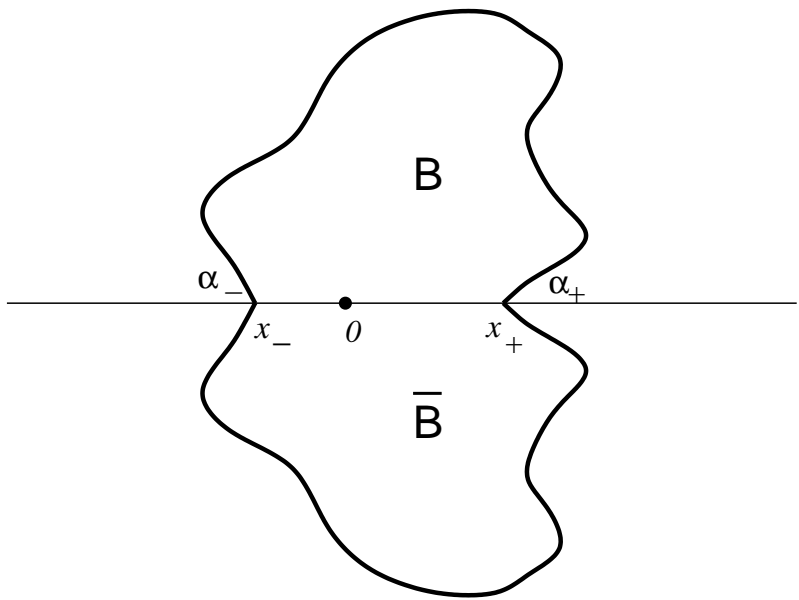}
        \caption{\it A fat slit $\BB$
        in the upper half plane and the
        corresponding symmetric domain $\DD = \BB \cup
\bar \BB$.}
    \label{fig:symdom}
\end{figure}

It is often convenient to treat fat slits as upper
halves of domains symmetric with respect to the real
axis. Namely,
set $\DD = \BB \cup \bar \BB$, where
$\bar \BB$ is the domain in the lower half plane
which is obtained from $\BB$ by complex conjugation
$z \to \bar z$ (Fig. \ref{fig:symdom}).
Obviously, the domain $\DD$ is symmetric
with respect to the complex conjugation.
In what follows we call such domains simply
{\it symmetric}. Vice versa, any compact simply connected
symmetric domain
$\DD$ is a union of a fat slit and its complex conjugate.
The boundary of $\DD$ is assumed to be analytic
everywhere except the two points on the real axis which are
allowed to be corner points.

\subsection{Conformal maps}

\begin{figure}[t]
   \centering
        \includegraphics[angle=-00,scale=0.45]{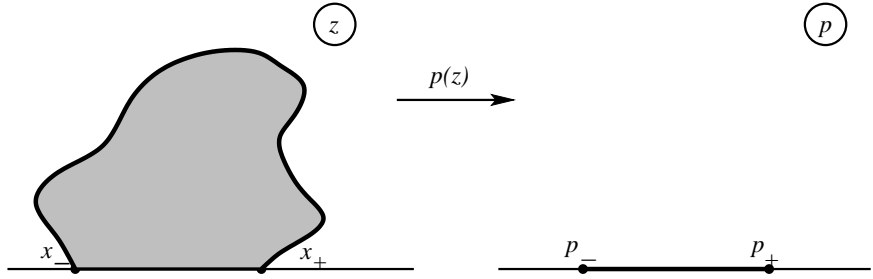}
        \caption{\it The conformal map $p(z)$.}
    \label{fig:map}
\end{figure}

Let $p(z)$ be a conformal map from $\HH \setminus \BB$
(in the $z$-plane) onto
$\HH$ (in the $p$-plane) shown schematically in Fig. \ref{fig:map}.
We normalize it by the condition that the expansion of
$p(z)$ in a Laurent
series at infinity is of the form
\beq\label{G40}
p(z)=z + \frac{u}{z}+\sum_{k\geq 2}u_k z^{-k}\,,
\quad |z| \to \infty
\eeq
(a ``hydrodynamic" normalization). Assuming this normalization,
the map is unique. The upper part of the boundary, $\gamma$,
is mapped to a segment of the real axis $[p_{-} , \, p_{+} ]$,
while the rays of the real axis outside $\BB$ are mapped
to the real rays $[-\infty , \, p_{-} ]$ and
$[p_{+} , \, \infty ]$ (Fig. \ref{fig:map}). From this it follows that
the coefficients $u_k$
are all real numbers.
The first coefficient, $u_1 :=u$, is called {\it a capacity}
of $\BB$. It is known to be positive.
We also need the inverse map,
$z(p)$, which can be expanded into the inverse Laurent series
\beq\label{G41}
z(p)=p - \frac{u}{p} +\sum_{k=2}^{\infty}a_k p^{-k}\,,
\quad |p| \to \infty
\eeq
with real coefficients $a_k$ connected with $u_k$ by
polynomial relations.
The series
converges for large enough $|p|$.

According to the Schwarz symmetry principle, the function
$z(p)$ admits an analytic continuation to the lower half plane.
This analytically continued
function performs a conformal map from the whole
complex plane with a
finite cut on the real axis from $p_{-}$ to $p_{+} $
onto the exterior of the symmetric domain
$\DD =\BB \cup \bar \BB$ .

\subsection{Green's functions}

Let $\DD =\BB \cup \bar \BB$ be a symmetric domain and
let $G(z,z')$ be the standard Green's function of the Dirichlet
boundary problem in $\CC \setminus \DD$.
The function is harmonic in $\CC \setminus \DD$
with respect to both
variables except at $z=z'$, where it has a logarithmic
singularity $G(z,z')=\log |z-z'|+\ldots$ and equals zero
when either $z$ or $z'$ lies on the boundary of $\DD$.
The Green's function solves the Dirichlet boundary
value problem: the formula
\beq\label{G1}
f(z)=-\, \frac{1}{2\pi}\oint_{\p \DD} f(\xi )\p_n G(z, \xi )
|d\xi |
\eeq
harmonically extends the function $f(\xi )$ from the contour
$\p \DD$ to its exterior. Here and below, $\p_n$ is the normal
derivative at the boundary, with the normal vector being directed
to the exterior of $\DD$.
We also note the
Hadamard formula \cite{Hadamard} for
variation of the Green's function under variation of the domain:
\beq\label{G2}
\delta G(a,b)=\frac{1}{2\pi}\oint_{\p \DD}\p_n G(a,z) \p_n G(b,z)
\delta n (z) |dz|\,,
\eeq
where $\delta n (z)$ is the infinitesimal normal displacement of the
contour. Some care is needed to define a deformation near the
corner points.
However, for our purposes it is enough to
consider deformations with fixed corner points, then
$\delta n (z)$ is well defined at any point of the boundary.
Also, in this paper we consider only the case when both angles
$\alpha_+$ and $\alpha_-$ are acute, $0 < \alpha_{\pm} <\pi /2$,
then the normal derivative of the Green's function vanishes at the corners
and the integral converges (see more details below).

For symmetric domains the Green's function obeys the property
$G(z,z')=G(\bar z, \bar z')$. Let us call a function $f(z)$
{\it even} (respectively, {\it odd}) if $f(z)=f(\bar z)$
(respectively, $f(z)=-f(\bar z)$). It is
natural to introduce even and odd Green's functions
such that $G^{\pm}(z, z')=\pm G^{\pm}(z, \bar z')$:
$$
G^{\pm}(z, z')=G(z, z')\pm G(z, \bar z')\,.
$$
Note that $G^{-}(x,z)=0$ for real $x$.
The Poisson formula (\ref{G1}) for even and odd boundary
functions can be written in the form
\beq\label{G3}
f(z)=-\, \frac{1}{2\pi}\int_{\gamma} f(\xi )\p_n G^{\pm}(z, \xi )
|d\xi |\,,
\eeq
where the integration goes over the
{\it non-closed} contour $\gamma$ (Fig. \ref{fig:db}).
The Hadamard formula for $G^{\pm}$,
\beq\label{G2a}
\delta G^{\pm}(a,b)=\frac{1}{2\pi}\int_{\gamma}\p_n G^{\pm}(a,z)
\p_n G^{\pm}(b,z)
\delta n (z) |dz|
\eeq
directly follows from (\ref{G2}), taking into account
that deformations of symmetric domains obey the condition
$\delta n (z)=\delta n (\bar z)$.

\subsection{The odd Green's function}

An important part in
what follows is played by the odd Green's
function $G^{-}$. It solves the following Dirichlet
boundary value problem in $\HH$: To find a harmonic function in
$\HH \setminus \BB$ bounded at infinity such that it is equal to
a given function on $\gamma$ and $0$ on the rays of the real
axis outside $\BB$.
Similar to the Green's function $G$, $G^-$ can be
expressed through a conformal map to a fixed reference domain.
The most natural reference domain in our case
is the upper half plane $\HH$.
It is easy to see that
\beq\label{G4}
G^{-}(z,z')=\log \left | \frac{p(z)-p(z')}{p(z)-p(\bar z')}\right |\,,
\eeq
where $p(z)$ is the conformal map (\ref{G40})
from $\HH \setminus \BB$ onto
$\HH$.
We also need a useful formula for the kernel
$\p_n G^- (a,z)$ in (\ref{G3}) through the
conformal map,
\beq\label{G43}
\p_n G^- (a,z)=-2\, {\cal I}m \, p(a)\,
\frac{|p'(z)|}{|p(z)-p(a)|^2}\,,
\quad z\in \gamma
\eeq
(which straightforwardly follows from (\ref{G4}))
and its limiting case as $|a|\to \infty$:
\beq\label{G42}
\p_n {\cal I}m \, p(z)=|p'(z)|\,, \quad z\in \gamma \,.
\eeq

Let us present the expansion of the odd Green's function
$G^- (a,z)$ as $|a|\to \infty$:
\beq\label{G5}
G^- (a,z)=2 \sum_{k\geq 1}\frac{1}{k}\, {\cal I}m (a^{-k})\,
{\cal I}m (B_k (p(z)))\,.
\eeq
Here $B_k (p)$ are Faber polynomials of $p(z)$ defined by
the expansion
\beq\label{G6a}
\log \frac{z}{p(z)-p} =\sum_{k\geq 1}\frac{z^{-k}}{k}B_k (p)\,,
\quad |z|\to \infty \,,
\eeq
and explicitly given by
\beq\label{G6}
B_k (p)=(z^k (p))_{\geq 0}\,,
\eeq
where $(\ldots )_{\geq 0}$ means the polynomial part of the
Laurent series.
Indeed, fixing a point $z_1 \in \HH \setminus \BB$, we
have
$$
\sum_{k\geq 1} \frac{z_{1}^{-k}}{k}\, B_k(p)=
\sum_{k\geq 1}\frac{\Bigl (z^k(p)\Bigr )_{\geq 0}}{kz^k (p_1)}
=-\left [ \log \left (1-\frac{z(p)}{z(p_1)}\right )\right ]_{\geq 0}
$$
where $p_1=p(z_1)$. To separate non-negative part, we write
$$
\log \left (1-\frac{z(p)}{z(p_1)}\right )=\log
\frac{p_1-p}{z(p_1)}+\log \frac{z(p_1)-z(p)}{p_1-p}
$$
and notice that the expansion of the first (second)
term contains only non-negative (respectively,
negative) powers of $p$.
Therefore,
$$
\sum_{k\geq 1}\frac{z_{1}^{-k}}{k}\, B_k(p)=-\log (p(z_1)-p)+
\log z_1,
$$
which coincides with (\ref{G6a}).
In particular, $B_1 (p)=p$. Clearly,
\beq\label{G7}
B_k (p(z))=z^k + O(z^{-1})\,, \quad |z|\to \infty \,,
\eeq
and the function $B_k (p(z))-z^k$ is analytic in $\CC \setminus \DD$.

\section{Local coordinates in the space of fat slits}

We are going to show that the harmonic moments $T_k$
(defined as in (\ref{H4}) below)
locally characterize a fat slit in the following sense.
First, any small deformation of $\BB$ that preserves
the moments $T_k$, is trivial, i.e.,
any non-trivial deformation changes at least
one of them. This fact means local uniqueness
of a fat slit having given moments. Second, the moments
$T_k$, under certain
conditions discussed below,
are independent quantities
meaning that one can
explicitly define infinitesimal
deformations of $\BB$ that change any one of them keeping
all other fixed. In this weak sense they serve as local
coordinates in the space of fat slits (c.f. the remark
in Section 2.1 in \cite{KMZ}).

\subsection{Harmonic moments}

Given a fat slit $\BB$, let us introduce harmonic
moments of the fat slit domain $\HH \setminus \BB$
as
\beq\label{H4}
\begin{array}{l}
\displaystyle{
T_k = \frac{2}{\pi k}\, {\cal I}m \int_{\HHH \setminus \BB}
z^{-k} d^2 z\,, \quad k\geq 2},\\ \\
\displaystyle{
T_1 =-\, \frac{2}{\pi }\, {\cal I}m \int_{\BB}z^{-1}d^2z}\,.
\end{array}
\eeq
Here we assume that the base of $\BB$ is a segment containing zero.
Since ${\cal I}m (z^{-1})<0$ for $z\in \HH$, $T_1$ is always
positive. Although the integrand
in the formula for $T_1$ is singular
at the origin, the integral converges.
The integral for $T_2$ diverges at infinity, so
one should introduce a cut-off at some large radius and make
the angular integration first; this prescription is equivalent to
the contour integral representation given below.
Note that this set of moments
does not include the area of $\BB$. Note also that the
standard harmonic moments dealt with in \cite{MWZ} are,
for symmetric domains, real parts of the integrals
in (\ref{H4}) rather than imaginary ones.

\begin{figure}[t]
   \centering
        \includegraphics[angle=-00,scale=0.5]{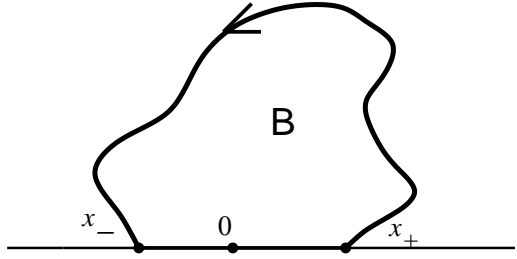}
        \caption{\it The integration contour $\gamma $.}
    \label{fig:db}
\end{figure}

Some other integral representations of the moments
(\ref{H4}) are also useful.
Imaginary parts of the integrals
can be taken by extending
the integration to the lower half plane as
\beq\label{H4a}
\begin{array}{l}
\displaystyle{
T_k =\frac{1}{i\pi k}\, \int_{\CCC \setminus \DD}
\mbox{sign} (y) \, z^{-k} d^2 z\,, \quad k\geq 2},\\ \\
\displaystyle{
T_1 =-\, \frac{1}{i\pi }\, \int_{\DD}\mbox{sign} (y) \,z^{-1}d^2z}\,,
\end{array}
\eeq
where $y={\cal I}m \, z$.
Contour integral representations are easily obtained
using the Stokes theorem. They read
\beq\label{H5}
T_k =\frac{2}{\pi k}\, {\cal I}m \int_{\gamma}yz^{-k}\, dz=
\frac{1}{\pi i k}\, \oint_{\p \DD}
|y|\, z^{-k}dz\,,
\quad k \geq 1\,.
\eeq
The non-closed integration contour $\gamma$ (shown in Fig.
\ref{fig:db}) is the part of the boundary of $\BB$ lying
in the upper half plane (with the orientation from right
to left).

It is convenient to introduce the generating function of the
moments $T_k$:
\beq\label{H6}
M_+ (z)=\frac{1}{\pi i}\oint_{\p \DD}\frac{|y'|\, dz'}{z'-z}
=\sum_{k=1}^{\infty} kT_k z^{k-1}\,, \quad |z|\to 0
\eeq
(here $y' ={\cal I}m \, z'$).
The integral of Cauchy type in (\ref{H6}) defines
an analytic function everywhere inside $\DD$. In a small
enough neighborhood of the origin this function is
represented by the (convergent) Taylor series standing
in the r.h.s. of (\ref{H6}).

For example, let $\BB$ be the half-disk of radius $R$:
$|z|\leq R$, ${\cal I}m \, z \geq 0$. Then an easy
calculation gives
$$
T_k = \frac{4 R^{2-k}}{\pi k^2 (2\! -\! k)} \quad
\mbox{for odd $k$}\,, \quad
\mbox{$T_k =0\; $ for even $k$}
$$
and
$$
M_+(z)=\frac{R}{\pi}\left (
2+ \frac{z^2 -R^2}{Rz}\log \frac{R-z}{R+z}\right ).
$$

\subsection{An electrostatic interpretation}

Similar to the standard harmonic moments from
papers \cite{WZ,MWZ},
the moments $T_k$ have a clear 2D electrostatic
interpretation. Let the interior of the domain
$\BB$ be filled by an electric charge with
uniform density $-1$ and let the lower half plane
(or just the real axis) be
a grounded conductor. By the reflection principle,
the electric potential $\Phi^- (z)$ in the upper
half plane is equal to the potential created by
the charge in $\BB$ and the fictitious ``mirror"
charge of opposite sign in $\bar \BB$:
\beq\label{E1}
\Phi^- (z)=-\, \frac{2}{\pi}\int_{\BB}
\log \left |\frac{z-z'}{z-\bar z'}\right | d^2z' \,.
\eeq
Let us show that the $T_k$'s are coefficients
in the multipole expansion of $\Phi^- (z)$ in the
interior of $\BB$ near the
origin. We have, for $z\in \BB$:
$$
\begin{array}{lll}
\p_z \Phi^- (z)&=&\displaystyle{
\frac{1}{\pi}\int_{\BB}\frac{d^2 z'}{z'-z}
-\, \frac{1}{\pi}\int_{\bar \BB}\frac{d^2 z'}{z'-z}}
\\ && \\
&=&-\bar z +\displaystyle{
\frac{1}{2\pi i}\oint_{\p \BB}\frac{\bar z' dz'}{z'-z}
-\, \frac{1}{2\pi i}\oint_{\p \bar \BB}\frac{\bar z' dz'}{z'-z}}
\\ && \\
&=&z-\bar z -\displaystyle{
\frac{1}{\pi }\oint_{\p \BB}\frac{y' dz'}{z'-z}
+ \frac{1}{\pi }\oint_{\p \bar \BB}\frac{y' dz'}{z'-z}}\,,
\end{array}
$$
where we substituted $\bar z'=z'-2iy'$ in the second line
and integrated the analytic parts by taking residues.
Since the function under the integrals vanishes on the
parts of the contours along the
real axis, we can eliminate them and combine the two integrals
into a single integral over $\p (\BB \cup \bar \BB)=\p \DD$:
\beq\label{E2}
\p_z \Phi^- (z)=z-\bar z -\,
\frac{1}{\pi }\oint_{\p \DD}\frac{|y'| dz'}{z'-z}\, = \,
z-\bar z -i\sum_{k\geq 1}kT_k z^{k-1}\,,
\eeq
where the second equality follows from (\ref{H6}). Since
$\Phi^- (x)=0$ at real $x$, we obtain the expansion of
the $\Phi^- (z)$ around $0$ in the upper half plane:
\beq\label{E3}
\Phi^- (z)=\frac{1}{2}(z-\bar z)^2 -i\sum_{k\geq 1}T_k
(z^k -\bar z^k) \,.
\eeq
Similarly, expanding $\Phi^-(z)$ around $\infty$ in the
upper half plane, we get:
\beq\label{E4}
\p_z \Phi^- (z)=-i\sum_{k\geq 1}V_k z^{-k-1}, \quad
\Phi^- (z)=i\sum_{k\geq 1}\frac{V_k}{k}(z^{-k}-\bar z^{-k})\,,
\eeq
where
\beq\label{E5}
V_k = \frac{2}{\pi}\, {\cal I}m  \int_{\BB}z^k \, d^2 z
\eeq
are moments of the interior. Their generating function, $M_-(z)$,
is given by the same
Cauchy-type integral (\ref{H6})
for $z$ outside $\DD$:
\beq\label{H6a}
M_- (z)=\frac{1}{\pi i}\oint_{\p \DD}\frac{|y'|\, dz'}{z'-z}
=\sum_{k=1}^{\infty} V_k z^{-k-1}\,, \quad |z|\to \infty \,.
\eeq

Sometimes an equivalent electrostatic interpretation
appears to be more convenient. Let us assume
that there is no conductor but the domain $\bar \BB$ in the lower
half plane is indeed filled with the ``mirror" charge.
In this case formulas (\ref{E2}), (\ref{E3}) and (\ref{E4})
admit continuation to the lower half plane which is achieved by
complex conjugation of both sides. For example, complex
conjugation of (\ref{E2}) yields
$\p_{\bar z} \Phi^- (\bar z)=\bar z- z +i\sum_{k\geq 1}kT_k \bar z^{k-1}$,
$z\in \BB$, which can be rewritten as
$\p_{z} \Phi^- (z)=z- \bar z +i\sum_{k\geq 1}kT_k z^{k-1}$,
$z\in \bar \BB$.

\subsection{Local uniqueness
of a fat slit with given moments}

Here, we prove the local
uniqueness
of a fat slit with given moments. The deformations
changing only one moment will be constructed
in the next subsection.

For the purpose of this section
it is convenient to work with symmetric domains
$\DD = \BB \cup \bar \BB$ rather than with fat slits themselves.
Let $\DD (t)$ be a one-parameter deformation in the class
of symmetric domains such that $\DD (0)=\DD$ and $\p_t T_k =0$
for all $k=1, 2, \ldots$. We shall show that any such deformation
is trivial: $\DD(t)=\DD (0)$ (at least in a small neighborhood of
$t=0$). To see this, consider the $t$-derivative of the function
$M_+(z)$. A simple calculation (see Appendix A) shows that
\beq\label{H7}
\p_t M_+(z)=\frac{1}{\pi i}\oint_{\p \DD}
\frac{\mbox{sign}(y')v_n (z')}{z-z'}\, |dz'|\,,
\eeq
where $v_n (z)=\delta n(z)/\delta t$ is the ``velocity" of
the normal displacement of the boundary at the point $z$.
(If $x(\sigma , t)$, $y(\sigma , t)$ is any parametrization
of the contour, then
$
v_n = \frac{d\sigma}{dl}(\p_{\sigma} y \p_t x - \p_{\sigma} x \p_t y)
$,
where $dl =|dz| =\sqrt{(dx)^2 +(dy)^2}$ is the line element along the
contour and $v_n$ is positive when the contour moves to the right
of the increasing $\sigma$ direction).
The Cauchy-type integral in the r.h.s. defines analytic
functions both inside and outside the contour. For $z$ inside the contour,
choosing a neighborhood of $0$ such that $|z|<|z'|$ for all
$z' \in \p \DD$, we can expand $M_+(z)$ as in (\ref{H6}) and find that
$$
\p_t M_+(z)=\sum_{k=1}^{\infty}k\p_t T_k z^{k-1} =0
$$
for all $z$ in this neighborhood. By uniqueness of analytic
continuation $\p_t M_+(z)=0$ everywhere in $\DD$.
According to the property of integrals of Cauchy type
this means that the function
$\p_t M_+(z)$ is analytic in $\CC \setminus \DD$ and
is given there by the Cauchy integral (\ref{H7}).
The boundary value of this function is
$sign\, (y)\, v_n(z)|dz|/dz$ with {\it real}
$v_n(z)$
almost everywhere on the contour (in our case actually everywhere
except maybe the two corner points on the real axis).
Furthermore, deformations
of symmetric domains preserving symmetry with respect to the
real axis obey the condition
$$
\oint_{\p \DD}\mbox{sign}(y)v_n (z)|dz| =0
$$
(because $v_n (z)=v_n (\bar z)$), which means,
according to the Cauchy integral representation, that
the function $\p_t M_+(z)$ has a zero
at infinity of at least second order. Invoking
the technique from the theory of boundary values
of analytic functions \cite{Privalov,Duren}, one can prove
(see Appendix B) that
any analytic function with such properties in $\CC \setminus \DD$
must be identically zero.
Therefore,
$v_n \equiv 0$, i.e., the deformation
is trivial.

\subsection{Special deformations of fat slits}

In order to define deformations that change one of the
moments $T_k$ keeping all other fixed we need the
odd Green's function for symmetric domains introduced
in Section 2.2.

Fix a point $a\in \HH \setminus \BB$ and consider
small deformations $\delta_{a}^{-}$ of
a fat slit $\BB$ defined by the
infinitesimal normal displacement of the boundary as follows:
\beq\label{D3}
\delta_{a}^{-} n (\xi )=
-\frac{\epsilon}{2}\p_n G^{-}(a, \xi )\,,
\quad \xi \in \gamma \,, \;\; a\in \HH \setminus \BB \,.
\eeq
(here $\xi \in \p \DD$ and $\epsilon \to 0$).
These are analogs of the deformations
$\delta_a n (\xi )=
-\frac{\epsilon}{2}\p_n G(a, \xi )$ from \cite{KMZ} which
generate dToda flows in the space of compact domains in the plane.
As we shall soon see, $\delta_{a}^{-}$ generate, in the
same sense, dKP-flows
in the space of fat slits.
Extending the definition
of the $\delta_{a}^{-}$ to the lower half plane, we can
define the corresponding deformation of the
symmetric domain $\DD =\BB \cup \bar \BB$:
\beq\label{D2}
\delta_{a}^{-} n (\xi )=-\frac{\epsilon}{2}\, \mbox{sign}\,
({\cal I}m \xi )\,
\p_n G^{-}(a, \xi )\,, \quad \xi \in \p \DD \,.
\eeq
Clearly, $\delta_{a}^{-}n (\xi )/\epsilon$ as $\epsilon \to 0$
is to be understood as normal velocity of the boundary
under the deformation.

An important comment is in order. For the deformations
$\delta_{a}^{-}$ to be well defined around the points
$x_{-}$, $x_{+}$, we assume that the angles $\alpha_{\pm}$
between the curve $\gamma$ and the real axis are strictly acute:
$0<\alpha_{\pm}<\pi /2$ (see Fig. \ref{fig:symdom}).
Since $p'(z)\sim (z-x_{\pm})^{\frac{\pi}{\alpha_{\pm}}-1}$
around the corner points (for a rigorous proof
see, e.g., \cite[Lemma 2.8]{AKS}), it is seen from (\ref{G43}) that
the normal velocity of the
boundary near the corner points tends to zero
as $\xi \to x_{\pm}$ and, moreover, so does the angular velocity
of the parts of the boundary near the corners (which is of order
$|p''(z)|$).
This means that the points $x_{\pm}$ and the angles
$\alpha_{\pm}$ remain fixed.
For not strictly acute angles $\alpha_+$, $\alpha_-$
the situation is much
more complicated. For example,
the angles can immediately
jump to other values and the deformations are not always
well defined (cf. \cite{KLV}). This case deserves further
investigation.

Expanding the Green's function as in (\ref{G5}), one can introduce
the deformations
\beq\label{D4}
\delta ^{(k, -)} n \, (\xi )=
\frac{\epsilon}{2}\,
\mbox{sign}\,
({\cal I}m \xi )\,
\p_n {\cal I}m B_k (p(\xi ) )\,,
\quad \xi \in \p \DD \,.
\eeq
Like $\delta_{a}^{-}$, the deformations
$\delta ^{(k, -)}$ do not shift the endpoints of $\gamma$.

It is not difficult to show that $\delta ^{(k, -)}$
changes the harmonic moment $T_k$ keeping all other fixed.
Indeed, assuming that $a \in \DD$, we write:
$$
\delta ^{(k, -)}M_+(a)=
\frac{1}{\pi i}\oint_{\p \DD}
\frac{\mbox{sign} (y)\delta ^{(k, -)} n \, (z)}{a-z}\, |dz|
=\frac{\epsilon}{2\pi i}\oint_{\p \DD}
\frac{\p_n {\cal I}m B_k (p(z))}{a-z}\, |dz|\,,
$$
where the extra sign in the definition
of $\delta ^{(k, -)}$ cancels
the $\mbox{sign} (y)$ in the definition of $M_+(a)$.
Because ${\cal I}m B_k (p(z))=0$ on $\p \DD$, we have:
$$
\delta ^{(k, -)}M_+(a)=
\frac{\epsilon}{2\pi i}\oint_{\p \DD}
\frac{dB_{k}(p(z))}{z-a}=\frac{\epsilon}{2\pi i}\oint_{\p \DD}
\frac{dz^{k}}{z-a}=\epsilon k a^{k-1}\,,
$$
where we have used the fact that the function $B_k -z^k$ is
analytic in $\CC \setminus \DD$ and vanishes at $\infty$,
and so does not contribute to the integral.
We thus see that $\delta ^{(k, -)}T_j = \epsilon \delta_{kj}$.

\subsection{Vector fields in the space of fat slits}

Deformations which depend on $\gamma$ in a smooth way
can be represented by vector fields in the space of fat
slits.
Let $\delta n (z)$ be any small deformation of a fat slit.
Given a functional $X$ on the space of fat slits, its
variation reads:
$$
\delta X = \int_{\gamma}\frac{\delta X}{\delta n (\xi )}
\, \delta n (\xi ) |d \xi |\,.
$$
The variational derivative
$\delta X/ \delta n (\xi )$ has the following meaning:
$$
\frac{\delta X}{\delta n (\xi )}=\lim_{\varepsilon \to 0}
\frac{\delta^{(\varepsilon )}X}{\varepsilon}\,.
$$
Here $\delta^{(\varepsilon )}X$ is the variation of the
functional under attaching  a small bump of area $\varepsilon \to 0$
at the point $\xi \in \gamma$ (a symmetric bump is assumed to
be attached at the point $\bar \xi$).
Let $\delta n(\xi ) = \epsilon g(\xi )$, $\epsilon \to 0$,
then we define the vector field (the Lie derivative)
$\nabla^{(g)}$:
$$
\nabla^{(g)}X=\int_{ \gamma}\frac{\delta X}{\delta n (\xi )}
\, g(\xi  )|d\xi |\,.
$$

Applying these general formulas to the deformations
$\delta_{a}^{-}$ (see (\ref{D3})), we can write
$\delta_{a}^{-}X = \epsilon \nabla^- (a)X$, where
the vector field
$\nabla^{-} (a)$ acts on functionals as follows:
\beq\label{V1}
\nabla^{-} (a)X=-\frac{1}{2}\int_{\gamma}
\frac{\delta X}{\delta n (\xi )}
\, \p_n G^{-}(a,\xi )|d\xi |\,.
\eeq
This equation gives an invariant definition of the
vector field $\nabla^- (a)$ independent of any choice
of coordinates.
According to the Dirichlet
formula (\ref{G3}), the action of $\nabla^- (a)$
provides the harmonic extension of the function
$\pi \delta X/\delta n (\xi )$ from $\gamma$ to $\HH$
bounded at infinity and equal to $0$ on the rays of the real
axis outside $\BB$.

In the local coordinates $T_k$, $\nabla^- (a)$ is
represented as an infinite
linear combination of the vector fields $\p / \p T_k$ which
can be thought of as partial derivatives.
To find it explicitly, we calculate
$$
\delta_{a}^{-}T_k = -\, \frac{2}{\pi k}\int_{\gamma}
{\cal I}m (z^{-k})\delta_{a}^{-}n(z)|dz| =
\frac{\epsilon}{\pi k}\int_{\gamma}
{\cal I}m (z^{-k})\p_n G^- (a,z)|dz|
\, = \,
-\, \frac{2\epsilon}{k}\, {\cal I}m (a^{-k}),
$$
where the last equality
follows from the Dirichlet formula (\ref{G3}) for symmetric
domains. Now, given a functional
$X$ on the space of fat slits, and assuming that
$X$ is a function of the moments $T_k$ only, we write
$$
\delta_{a}^{-}X =\sum_{k\geq 1}\frac{\p X}{\p T_k}\delta_{a}^{-} T_k =
-2\epsilon \sum_{k\geq 1}\frac{1}{k}\,
{\cal I}m (a^{-k})\frac{\p X}{\p T_k}
= \epsilon \nabla^- (a)X \,,
$$
so $\nabla^- (a)$ is given by
\beq\label{D5}
\nabla^- (a)=-2\sum_{k\geq 1}\frac{1}{k}\,
{\cal I}m (a^{-k})\frac{\p }{\p T_k}\,.
\eeq

The vector fields  $\nabla^- (a)$ are ``half-plane" analogs of the
vector fields $\nabla (a)$ introduced
in \cite{KMZ} via their action on functionals $X$ in the space of all
domains:
\beq\label{V111}
\nabla (a)X=-\frac{1}{2}\int_{\p \DD}
\frac{\delta X}{\delta n (\xi )}
\, \p_n G(a,\xi )|d\xi |
\eeq
(cf. (\ref{V1})).
Below we will show that
the dKP hierarchy is related to the vector fields
$\nabla^- (a)$ in the same way as the dToda hierarchy is related
to the $\nabla (a)$.

\section{The dispersionless KP hierarchy}

\subsection{Lax equations}

The dispersionless KP (dKP) hierarchy is encoded in
the Hadamard formula (\ref{G2a}). To see this, fix three
points $a,b,c \in \HH \setminus \BB$ and find $\delta_{c}^{-}G^- (a,b)$:
\beq\label{KP1}
\delta_{c}^{-}G^- (a,b)=-\, \frac{\epsilon}{4\pi}
\int_{\gamma} \p_n G^- (a,z)\p_n G^- (b,z)\p_n G^- (c,z)|dz|\,.
\eeq
This formula shows that the quantity $\nabla^- (a)G^- (b,c)$
is {\it symmetric} with respect to all three arguments:
\beq\label{KP2}
\nabla^- (a)G^- (b,c)=\nabla^- (b)G^- (a,c)=\nabla^- (c)G^- (a,b)\,.
\eeq
Using the expansions (\ref{G5}) and (\ref{D5}), we get
$$
\frac{\p}{\p T_k}{\cal I}m (B_l (p(z))=
\frac{\p}{\p T_l}{\cal I}m (B_k (p(z))\,,
\quad k,l \geq 1 \,.
$$
Since $\overline{B_k (p(z))}=B_k (p(\bar z))$, we can easily separate
holomorphic and antiholomorphic parts of this equality and rewrite
it as a relation between functions of $z$ only:
\beq\label{KP3}
\frac{\p B_l (p(z))}{\p T_k}=\frac{\p B_k (p(z))}{\p T_l}\,.
\eeq
In particular,
\beq\label{KP4}
\frac{\p B_k (p(z))}{\p T_1}=\frac{\p p(z)}{\p T_k}\,.
\eeq
Treating $p$ rather than $z$
as an independent variable and passing to the inverse map, $z(p)$,
one can bring this equality to the form
\beq\label{KP5}
\frac{\p z(p)}{\p T_k}=\frac{\p B_k(p)}{\p p}\frac{\p z(p)}{\p T_1}-
\frac{\p B_k(p)}{\p T_1}\frac{\p z(p)}{\p p}
\equiv \{B_k (p),\, z(p)\}\,.
\eeq
Recalling that $B_k (p)=(z^k(p))_{\geq 0}$ (see (\ref{G6})),
we recognize the standard Lax equations of the dKP hierarchy.

So far we assumed that $p$ does not belong to the segment
$[p_- , p_+ ]$ (see Fig. \ref{fig:map}).
This segment is a branch cut
of the function $z(p)$. On this cut we can write
\beq\label{height}
z(p\pm i0)=x(p)\pm i |y(p)|\,, \quad p_- < p < p_+ \,.
\eeq
Here $|y(p)|=y(p+i0)=-y(p-i0)\geq 0$.
Outside the cut,
the real-valued function $x(p)$ has the same expansion
(\ref{G41}) as the $z(p)$. On the cut, it is given by the principal
value integral
\beq\label{PV}
x(p)=p+\frac{1}{\pi}{\rm P.V.}\! \int_{p_-}^{p_+}
\frac{|y(p')| dp'}{p'-p}\,.
\eeq
Because $B_k(p)$ is a polynomial with real coefficients,
one sees from (\ref{KP5}) that $x(p)$ and $|y(p)|$ obey
the same Lax equations:
\beq\label{KP51}
\frac{\p x(p)}{\p T_k}=\{B_k (p), \, x(p)\}\,,
\quad
\frac{\p |y(p)|}{\p T_k}=\{B_k (p), \, |y(p)|\}\,,
\quad
p\in [p_- , p_+ ] \,.
\eeq
In the next subsection we
show that $2|y(p)|$ is the Orlov-Shulman function.

\subsection{The Orlov-Shulman function}

Consider the functions $M_+(z)$, $M_-(z)$ defined by
the integrals of Cauchy type (\ref{H6}), (\ref{H6a}) for $z$ inside
and outside the domain $\DD =\BB \cup \bar \BB$ respectively.
They can be also represented
by the Taylor series
$$
M_+(z)=\sum_{k\geq 1}kT_k z^{k-1}\,,
\quad
M_-(z)=\sum_{k\geq 1}V_k z^{-k-1}
$$
which converge in some neighborhoods of
$0$ and $\infty$ respectively. The function $M_+(z)$ is analytic
everywhere in $\DD$ while $M_-(z)$ is analytic everywhere in
$\CC \setminus \DD$ (with zero of second order at $\infty$).
Moreover, for analytic arcs $\gamma$ both $M_+(z)$ and
$M_-(z)$ can be
analytically continued across the arcs $\gamma$ and $\bar \gamma$
everywhere except
their endpoints on the real axis, where both functions
have a singularity. Therefore, the function
\beq\label{M1}
M(z):=M_+(z)-M_-(z)=\sum_{k\geq 1}kT_k z^{k-1}-\sum_{k\geq 1}V_k z^{-k-1}
\eeq
is analytic in a neighborhood of the boundary of
$\DD$ (excluding the points $x_{\pm}$ on the real axis)
and, by the property of the Cauchy-type integrals,
is equal to $2 |{\cal I}m z| =2|y|$ on $\gamma \cup \bar \gamma$.
This can be also seen from formulas (\ref{E2}), (\ref{E4})
(together with their extensions to the lower half plane) taking into
account that the derivatives of the electrostatic potential
are continuous at the boundary:
$$
\p_z \Phi^- (z)\Bigr |_{{\rm in}}=2iy \mp iM_+(z)\, \, =\, \,
\mp iM_-(z)=
\p_z \Phi^- (z)\Bigr |_{{\rm out}}
$$
where the upper (lower) sign is taken for $z$ in the
upper (lower) half plane.
We see that for $z$ in the upper half plane
$M(z)$ is the analytic continuation of the function $2y$ from
the contour $\gamma$, while for $z$ in the lower half plane
$M(z)$ is the analytic continuation of the function $-2y$ from
the complex conjugate contour $\bar \gamma$, i.e.,
$M(z(p))=2|y(p)|$.

Let $S(z)$ be the Schwarz function of
the contour $\gamma$, i.e., an analytic function
such that $S(z)=\bar z$ for $z$ on $\gamma$ (see
\cite{Davis} for details). For analytic contours, it is known to be
well defined in some strip-like neighborhood of the curve.
Clearly, the Schwarz of the complex conjugate contour $\bar \gamma$
is then $\bar S(z)=\overline{S(\bar z)}$.
By uniqueness of analytic continuation, we can express $M(z)$
in terms of the Schwarz function:
\beq\label{M2}
M(z)=\left \{
\begin{array}{l} \,\, i(S(z)\, - \, z), \quad {\cal}Im \, z >0
\\ \\ -i (\bar S(z)-z), \quad {\cal}Im \, z <0\,.
\end{array}\right.
\eeq

Let us show that
\beq\label{M3}
\p_{T_k}M(z)=\p_zB_k(p(z))\,.
\eeq
Consider the change of
the $S(z)$ under the deformation
$\delta_{a}^{-}$. If $z\in \gamma$, then, using the
identity $\displaystyle{v_n (z)=\frac{\p_T S(z)}{2i \sqrt{S'(z)}}}$
for the normal velocity of the boundary under a deformation
with a parameter $T$, we can write
$$
\delta_{a}^{-} S(z)=-2i \sqrt{S'(z)}\, \frac{\epsilon}{2}
\, \p_n G^- (a,z)\,.
$$
Since $\sqrt{S'(z)} =|dz|/dz =1/\tau (z)$, where $\tau (z)$
is the unit tangent vector to the curve $\gamma$ (represented
as a complex number) and $\p_n G^- (a,z)=-2i\tau (z)\p_z G^- (a,z)$,
we have
$$
\delta_{a}^{-} S(z)=-2\epsilon \p_z G^- (a,z)
$$
and so,
\beq\label{M4}
\nabla^- (a)S(z)=-2\p_z G^- (a,z)
\eeq
for $z\in \gamma$ and, by analytic continuation,
everywhere in the neighborhood where $S(z)$ is
a well defined analytic function. Expanding both sides
as in (\ref{G5}), (\ref{D5}), we finally get
\beq\label{M5}
\p_{T_k}S(z)=-i \, \p_z B_k(p(z)),
\eeq
which is equivalent to (\ref{M3}).

An important particular case of (\ref{M3}) is
\beq\label{M6}
\p_{T_1}M(z)=\p_z p(z)\,.
\eeq
Passing to partial derivatives at constant $p$,
one can rewrite it in the form of the ``string equation":
\beq\label{M7}
\{z(p), \, 2|y(p)|\}=1\,, \quad p\in [p_- , p_+ ] \,.
\eeq
This relation together with the Lax equations (\ref{KP51})
show that $M(z(p))=2|y(p)|$ is the Orlov-Shulman function \cite{OS}
of the dKP hierarchy which describes deformations of fat slits.

A closely related useful object is the indefinite integral
of $M(z)$.
Using the notation of the previous subsection,
we introduce the function
\beq\label{Omega1}
\begin{array}{lll}
\Omega (z)&=&\displaystyle{\int_{0}^{z}M_+(z)dz +
\int_{z}^{\infty}M_-(z)dz}
\\&&\\
&=&\displaystyle{\sum_{k\geq 1}T_k z^k +\sum_{k\geq 1}
\frac{V_k}{k}\, z^{-k}}\,.
\end{array}
\eeq
It is analytic in the same strip-like neighborhood
of the curve $\gamma$ where the Schwarz function is
well defined. Since the electrostatic potential
$\Phi^- (z)$ is continuous on $\gamma$, it follows from
(\ref{E3}), (\ref{E4}) that
\beq\label{Omega2}
{\cal I}m \, \Omega (z)=y^2\,, \quad z\in \gamma \,.
\eeq
The real part of $\Omega (z)$ at $z\in \gamma$ has the
meaning of the partial area beneath the curve $\gamma$.
More precisely, let
$$
A(z)=\int_{x_-}^{x}y'dx'\,, \quad z=x+iy \in \gamma
$$
be the partial area of $\BB$ cut from the right by
a line orthogonal to the real axis and passing through
$z$, then
\beq\label{Omega3}
{\cal R}e \, \Omega (z_1)-{\cal R}e \, \Omega (z_2)=
2(A(z_1)-A(z_2))\,, \quad z_1, z_2 \in \gamma \,.
\eeq
By construction,
partial $T_k$-derivatives of the function $\Omega$ at constant $z$
are the generators of the flows:
\beq\label{Omega4}
B_k(p(z))=\p_{T_k}\Omega (z)\,.
\eeq
In this sense the function $\Omega =\Omega (z; \{T_j\})$ solves
the whole set of equations (\ref{KP3}) and thus provides a solution
to the dKP hierarchy.

\subsection{The string equation}

The string equation can be also derived in a different way
along the lines of \cite{MWZ}. The idea is to use equation (\ref{KP2})
with one of the three points lying on the contour $\gamma$:
$$
\nabla^- (\xi )G^- (a,b)=\nabla^- (a )G^- (b,\xi )\,,
\quad \xi \in \gamma \,,
$$
and the other two points tending to infinity. From (\ref{V1})
we see that the l.h.s. is equal to $\frac{1}{2}\p_n G^- (a,\xi )
\p_n G^- (b, \xi )$ which is $2{\cal I}m (a^{-1}){\cal I}m (b^{-1})
(\p_n {\cal I}m \, p(\xi ))^2$ as $a,b \to \infty$. The r.h.s.
in the same limit is $-4{\cal I}m (a^{-1}){\cal I}m (b^{-1})
\p_{T_1}{\cal I}m \, p(\xi )$. Equating them, we obtain the
important relation
\beq\label{string1}
2\p_{T_1}{\cal I}m \, p(z)=-|\p_z p(z)|^2\,, \quad z\in \gamma \,.
\eeq
Its extension to the lower half of the boundary of
$\DD$ reads
\beq\label{string1a}
2\p_{T_1}{\cal I}m \, p(z)=|\p_z p(z)|^2\,, \quad z\in \bar \gamma \,.
\eeq
Passing to the variable $p$ with the help of the identity
$$
\p_{T_1}z(p)=-\, \frac{\p_{T_1}p(z)}{\p_z p(z)}
$$
we rewrite equations (\ref{string1}), (\ref{string1a}) in the form
\beq\label{string2}
\frac{\p z(p-i\epsilon)}{\p p}\frac{\p z(p+i\epsilon)}{\p T_1}-
\frac{\p z(p-i\epsilon)}{\p T_1}\frac{\p z(p+i\epsilon)}{\p p}=i\,
\mbox{sign}\, \epsilon \,,
\quad \epsilon \to 0 \,.
\eeq
Note that the equation at $\epsilon <0$ is obtained from the one
at $\epsilon >0$ by complex conjugation.
Passing to the functions $x(p)$
and $y(p)$, we see that (\ref{string2})
coincides with (\ref{M7}).

Eq. (\ref{string2}) can be cast into the form
of an evolution equation for $z(p)$. To derive it, we
start from the Hadamard formula for the deformation
$\delta^{(1,-)}$:
$$
\delta^{(1,-)}G^-(a,z)=\frac{\epsilon}{4\pi}
\int_{\gamma}\p_n G^-(a,\xi )\p_n G^-(z,\xi )\,
\p_n {\cal I}m \, p(\xi )|d\xi | \,.
$$
Expanding the Green's function at $|a|\to \infty$
(see (\ref{G5})) and using the fact that
$\delta^{(1,-)}T_1 =\epsilon$, $\delta^{(1,-)}T_k =0$
at $k\geq 2$, we rewrite it as
$$
\p_{T_1}{\cal I}m \, p(z) = \frac{1}{4\pi}
\int_{\gamma}\p_n G^-(z,\xi )\,
|p'(\xi )|^2 |d\xi |\,,
$$
which, after extracting the holomorphic part and passing
to the integration in the $p$-plane, reads
$$
\p_{T_1}p(z)=\frac{1}{2\pi}\int_{p_-}^{p_+}
\frac{dp}{(p(z)-p)|z'(p)|^2}\,.
$$
In terms of the function $z(p)$ we have:
\beq\label{string5}
\p_{T_1}z(p)=\frac{z'(p)}{2\pi}
\int_{p_-}^{p_+}\frac{dp'}{(p'-p)|z'(p')|^2}\,,
\eeq
which is the desired evolution equation equivalent to
(\ref{string2}). The latter is obtained from
(\ref{string5}) by taking the jump of both sides
across the segment $[p_- , p_+ ]$. Because
the function $z(p)$ is analytic in the upper half plane,
this is equivalent to the full equation (\ref{string5}).

\section{The dKP hierarchy in the Hirota form}

Here we reformulate the dKP hierarchy in the Hirota form
using the
dispersionless ``tau-function" (free energy) and clarify
the geometric meaning of the latter.

\subsection{Functional $F^{-}$ and its variations}

Given a domain $\DD$, not necessarily symmetric,
one can introduce
the functional $F$:
\beq\label{F1}
F=-\, \frac{1}{\pi^2}\int_{\DD}\int_{\DD}\log
|z^{-1}-\zeta^{-1}|\, d^2 z d^2 \zeta \,.
\eeq
If the boundary is analytic,
$F$ is the ``tau-function for analytic curves" introduced in \cite{KKMWZ}
(more precisely, a
properly rescaled logarithm of the dToda tau-function).
In the 2D electrostatic interpretation, it has the meaning of
the electrostatic energy of a uniformly charged
domain $D$ with a compensating
point-like charge at the origin.
It was found in \cite{KKMWZ} that the Green's function
$G(a,b)$ is given by
\beq\label{F4}
G(a,b)=\log |a^{-1}-b^{-1}|+\frac{1}{2}
\nabla (a)\nabla (b)F,
\eeq
where the vector field $\nabla (z)$ in the space of
all domains is defined in
(\ref{V111}) (see \cite{MWZ,KMZ} for more details).
We are going to derive an analog of equation (\ref{F4}) for
fat slits in the upper half plane.

Given a fat slit $\BB$,
consider the following functional:
\beq\label{F3}
F^-=-\, \frac{1}{\pi^2}\int_{\BB}\int_{\BB}\log \left |
\frac{z-\zeta}{z-\bar \zeta}\right |\, d^2 z d^2 \zeta \,.
\eeq
It has the meaning of the 2D electrostatic energy of the
uniformly charged fat slit in the presence of a conductor
placed along the real axis. Taking a variation of $F^-$,
it is easy to find how the vector field $\nabla^- (a)$
acts on $F^-$. We use the general relations given in
section 4.2. We have:
$$
\pi \, \frac{\delta F^-}{\delta n (\xi )}=-\,
\frac{2}{\pi}\int_{\BB}\log \left | \frac{z-\xi}{z-\bar \xi}
\right | d^2 z\,.
$$
The function in the r.h.s. is already harmonic and bounded
in $\HH \setminus \BB$ as it stands, hence
\beq\label{F6}
\nabla^- (a)F^- =-\,
\frac{2}{\pi}\int_{\BB}\log \left | \frac{a-z}{a-\bar z}
\right | d^2 z =  \Phi^- (a)\,.
\eeq
Expanding both sides as $a\to \infty$ and comparing
coefficients in front of the basis harmonic functions
${\cal I}m (a^{-k})$, we find:
\beq\label{F7}
\frac{\p F^-}{\p T_k}= \frac{2}{\pi}\,
{\cal I}m  \int_{\BB}z^k \, d^2z = V_k\,,
\eeq
where $V_k$
are the interior harmonic moments. Proceeding in a similar way,
we find
$$
\pi \, \frac{\delta \Phi^- (a)}{\delta n (\xi )}=-\, 2
\log \left |
\frac{\xi - a}{\xi - \bar a}\right |\,.
$$
The r.h.s. is harmonic (in $\xi$) everywhere in
$\HH \setminus \BB$ except the point $a$. This singularity
can be canceled by adding the Green's function $G^-$ (which
vanishes on the boundary). Therefore,
$$
\nabla^- (b)\Phi^- (a)=-2\log \left |
\frac{a-b}{a-\bar b}\right | +2G^- (a,b)\,,
$$
and we obtain
the formula for $G^- (a,b)$,
\beq\label{F5}
G^- (a,b)=\log \left |
\frac{a-b}{a-\bar b}\right | +\frac{1}{2}
\nabla^- (a)\nabla^- (b)F^- \,,
\eeq
which is a ``half-plane" analog of (\ref{F4}).

\subsection{Hirota equations for the dKP hierarchy}

Equation
(\ref{F4}) is known to encode the dToda hierarchy
in the Hirota form.
Equation (\ref{F5}) does the same for the dKP
hierarchy. To see this, let us apply the arguments from \cite{KMZ}.

Combining (\ref{F5}) and (\ref{G4}), we obtain the relation
\beq\label{HH1}
\log \left | \frac{p(z)-p(z')}{p(z)-p(\bar z')}\right |^2
=\log \left | \frac{z-z'}{z-\bar z'}\right |^2
+\nabla^- (z)\nabla^- (z')F^-,
\eeq
which implies an infinite hierarchy of differential
equations for the function $F^-$. Recall that the conformal
map $p(z)$ is normalized as
\beq\label{HH2}
p(z)=z+\frac{u}{z}+ O(1/z^2)\,, \quad z\to \infty
\eeq
(see (\ref{G40})).
Tending $z'\to \infty$ in (\ref{HH1}), one gets:
\beq\label{HH3}
{\cal I}m \, p(z) ={\cal I}m \, z +\frac{1}{2}\,
\p_{T_1}\nabla^- (z)F^- \,.
\eeq
The limit $z\to \infty$ of this equality yields a simple
formula for the capacity:
\beq\label{HH4}
u= \p ^{2}_{T_1}F^- \,.
\eeq
Let us separate holomorphic parts
of these equations, introducing the holomorphic
part of the operator $\nabla^- (z)$:
\beq\label{HH5}
D (z)=\sum_{k\geq 1}\frac{z^{-k}}{k}\, \p_{T_k}\,,
\quad
\nabla^- (z)=i\, [D(z)\! -\! D(\bar z)]\,.
\eeq
Equation (\ref{HH1}) then implies the
relation
\beq\label{HH1a}
\log \frac{p(z)-p(z')}{p(z)-p(\bar z')}
=\log \frac{z-z'}{z-\bar z'}
+i D(z)\nabla^- (z')F^- ,
\eeq
which is holomorphic in $z$. In the limit $z' \to \infty$
it gives the formula for the conformal map $p(z)$:
\beq\label{HH6}
p(z)=z+ \p_{T_1}D(z)F^-
\eeq
(this formula also follows from (\ref{HH3})). In a similar way,
equation (\ref{HH1a}) implies the relation
\beq\label{HH1b}
\log \frac{p(z)-p(z')}{z-z'}=
-D(z)D(z')F^- ,
\eeq
which is holomorphic in both $z$ and $z'$. Taking into account
(\ref{HH6}), we rewrite it as follows:
\beq\label{HH7}
1\, - \, e^{-D(a)D(b)F^-}=
-\, \frac{D(a)-D(b)}{a-b}
\p_{T_1}F^-.
\eeq
It is the dKP hierarchy in the Hirota form. We see that
the function $F^-$ is the dispersionless tau-function for
this hierarchy. The double integral
representation (\ref{F3}) clarifies its geometric
meaning.

\section{A growth model associated with dKP hierarchy}

The special deformations from section 4.1 suggest to introduce
a growth model which is associated with the dKP hierarchy in the same way
as the Laplacian growth \cite{list,book} of compact planar domains
at zero surface tension
is associated \cite{MWWZ} with the dToda hierarchy. In fact
the model to be introduced is also of the Laplacian type, i.e.,
the interface dynamics is governed by the Darcy law, but
differs from the standard one by boundary conditions.

The idea should be already clear from section 4.1: to consider
growth of a fat slit under the deformation $\delta^{(1,-)}$ which
changes only the first harmonic moment $T_1$ keeping all other
fixed and to identify $T_1$ with time $T$.

The corresponding growth problem can be formulated as follows.
Consider a fat slit $\BB (T)$ with a moving boundary $\gamma (T)$,
where $T$ is time, and suppose that the motion of the boundary follows
the Darcy law:
\beq\label{gr1}
v_n (\xi )=\frac{1}{2}\, \p_n \phi (\xi )\,, \quad \xi \in \gamma \,.
\eeq
Here $v_n (\xi )=\delta n(\xi )/\delta T$ is the normal velocity
of the boundary at the point $\xi$ and $\phi (z)$ is a harmonic function
in $\HH \setminus \BB$ such that
\begin{itemize}
\item[(i)]
$\phi =0$ on $\gamma$ and on the
rays of the real axis $[-\infty , x_- ]$, $[x_+ , +\infty ]$;
\item[(ii)]
$\phi (z) = {\cal I}m \, z +o(1)$ as ${\cal I}m \, z \to +\infty$.
\end{itemize}
Clearly, $\phi (z)={\cal I}m \, p(z)$, where $p(z)$ is the
conformal map (\ref{G40}), is harmonic in $\HH \setminus \BB$
and obeys these conditions. Comparing with (\ref{D4}) at $k=1$,
we see that the dynamics is given by the deformation $\delta^{(1,-)}$
at any point in time $T=T_1$ and all the higher moments $T_k$ are integrals
of motion. In other words, for our growth process $\p_T M_+(z)=1$.
Equivalently, the dynamics can be reformulated in terms of
the inverse conformal map $z(p,T)$ as the ``string equation" in the form
(\ref{string2}),
\beq\label{LGE}
{\cal I}m \Bigl [\p _p z(p-i0,T)\, \p _T z(p+i0,T)\Bigr ]=
\frac{1}{2}\,,
\quad p\in [p_- , p_+ ]\,,
\eeq
or in the form of the evolution equation (\ref{string5})
(a similar equation for the Laplacian growth in the standard
setting is well known \cite{BS}).
As was already mentioned, the growth process is well defined
if both angles $\alpha_{\pm}$ between
$\gamma$ and the real axis are acute.
Then these angles
and the points $x_-$, $x_+$ stay fixed all the time.

\begin{figure}[t]
   \centering
        \includegraphics[angle=-00,scale=0.45]{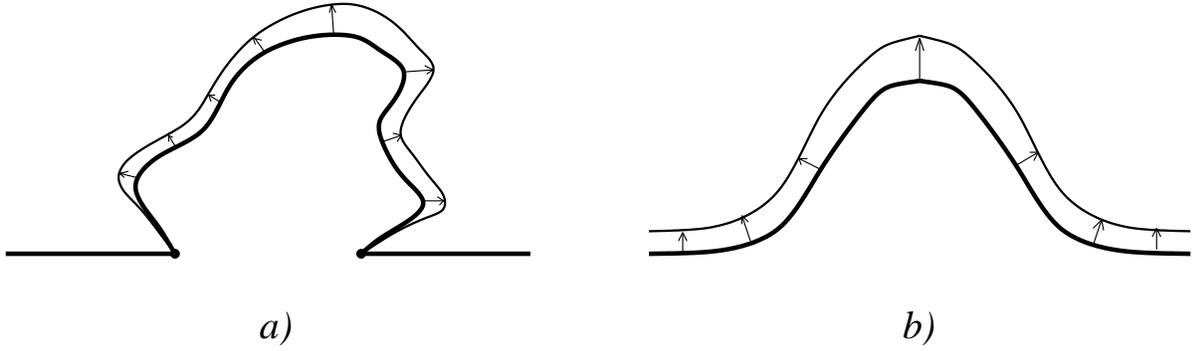}
        \caption{\it a): The Laplacian growth of a fat slit associated
        with dKP hierarchy;
        b) The standard Laplacian growth in the upper half plane.}
    \label{fig:lg}
\end{figure}

Comparing this setting with the standard Laplacian growth
in the upper half plane,
we see that the conditions on $\phi$ are very similar
if not the same: $\phi =0$
on an infinite contour from left to right
infinity, harmonic above it and tends to ${\cal I}m \, z$
as ${\cal I}m \, z\to +\infty$. However, in our case,
unlike in the standard one,
only a finite part of the boundary (namely, the part which lies above the
real axis) moves according to the Darcy law while the remaining part
(the rays of the real axis) is kept fixed despite the fact that the gradient of
$\phi$ is nonzero there (Fig. \ref{fig:lg}). We do not know a proper hydrodynamic
realization of this growth process.

So far we assumed that $x_-$ was strictly less than $x_+$,
so that the base of a fat slit was a segment of nonzero length.
The setting of this section allows us to consider the degenerate case
$x_-=x_+ =0$ when the base of a fat slit consists of one point. The
first harmonic moment $T_1 =T$ as well as the function
$M_+(z)$ are still well-defined but $M_+(z)$ is singular
at $z=0$ and thus can not be expanded into
the Taylor series around this point (this means
that the higher harmonic moments (\ref{H4})
are ill-defined).

\begin{figure}[t]
   \centering
        \includegraphics[angle=-00,scale=0.65]{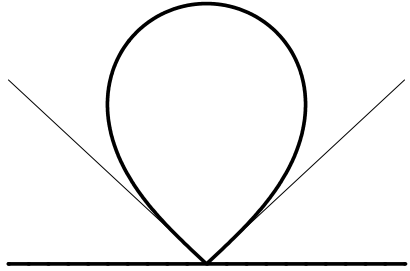}
        \caption{\it The curve $|z^2 +T^2|=T^2$ in the
        upper half plane. The tangent lines at the
        origin are at an angle of
        $45^0$ to the real axis.}
    \label{fig:example}
\end{figure}

The simplest explicit solution to equation (\ref{LGE}) known to us
describes self-similar growth of a ``fat slit" with degenerate base.
The function
\beq\label{ex1}
p(z,T)=\frac{z^2 +2T^2}{\sqrt{z^2 +T^2}}
\eeq
performs a conformal map from the exterior in $\HH$ of the curve
$|z^2 +T^2|=T^2$, or, in polar coordinates,
\beq\label{ex2}
R(\theta )=T\sqrt{-2\cos 2\theta}\,, \quad \frac{\pi}{4} \leq \theta
\leq \frac{3\pi}{4}\,,
\eeq
to the upper half plane. This curve is shown
in Fig. \ref{fig:example}. In this case
$x_- =x_+ =0$,
$\alpha _- = \alpha _+ =\pi /4$, $p_{\pm}=\pm 2T$.
The inverse map has the form
\beq\label{ex3}
z(p,T)=\frac{1}{\sqrt{2}}(p^2 -4T^2)^{1/4}
\Bigl (p+(p^2 -4T^2 )^{1/2}\Bigr )^{1/2}.
\eeq
One can check that it does solve equation (\ref{LGE}).
More results on explicit solutions to Laplacian growth of fat slits
will be published elsewhere.

\section{Concluding remarks}

We have constructed a parametric family of conformal maps
of the upper half plane which is related to the dKP hierarchy
with real ``times" $T_k$ in the same way as conformal maps of
the unit disk onto compact domains in the plane with smooth boundary
are related to the dToda hierarchy with complex conjugate ``times"
$t_k$, $\bar t_k$. Like in the dToda case, the deformations of
domains (``fat slits") in the upper half plane induced by dKP flows
have a physical interpretation as Laplacian growth with certain
type of sources or sinks at infinity. At the same time,
our construction extends the well known connection between the dKP hierarchy
and conformal maps of slit domains. In all cases, the conformal
map plays the role of the Lax function.

However,
a limiting procedure from fat slits
to usual slits is singular and is
not easy to trace on the level of the Lax equations.
We hope that a better understanding of this limit will further clarify
the geometric meaning of solutions to equations of the
dKP hierarchy. We also expect
that yet more general solutions can be obtained by the
same method applied to the case of a background
charge distributed in the upper half plane with a non-uniform density,
in accordance with a similar construction given in \cite{Z01}.

The solutions discussed in this paper have a nice geometric meaning
but it seems to be very hard to express them analytically in a closed
form. In this respect the situation is less favourable than in the
dToda case, where some simple explicit solutions for conformal maps
as functions of a finite number of nonzero harmonic moments
(corresponding, for example, to a parametric family of ellipses)
are available. It is clear that in our case the situation when
only a finite number of the moments $T_k$
are nonzero can not be realized because the local behavior
of their generating
function $M_+(z)$  near the points $x_-$, $x_+$
(which can be found from the integral representation
(\ref{H6})) shows that they are branch points
of this function. This suggests that the
corresponding solutions may be similar to the multi-cut
solutions to Laplacian growth discussed recently in \cite{AMWZ}.

\section{Appendices}

\subsection*{Appendix A}

Here we give some details of the derivation
of the formula (\ref{H7}) for the time derivative of the
function $M_+(z)$:
\beq\label{appA1}
\p_t M_+(a)=\frac{1}{\pi i}\oint_{\p \DD}
\frac{\mbox{sign}(y)v_n (z)}{a-z}\, |dz|\,.
\eeq
Here $v_n (z)$ is the velocity of
the normal displacement of the boundary at the point $z$.
Let $z(\sigma , t) =x(\sigma , t)+i y(\sigma , t)$
be any parametrization
of the contour such that $\sigma$ is a steadily increasing
function of the arc length, then it is a simple kinematical fact that
\beq \label{appA2}
v_n = \frac{d\sigma}{dl}(\p_{\sigma} y \p_t x - \p_{\sigma} x \p_t y)\,,
\eeq
where $dl =|dz| =\sqrt{(dx)^2 +(dy)^2}$ is the line element along the
contour. According to our convention,
$v_n (z)$ is positive when
the contour, in a neighborhood of the
point $z$, moves to the right
of the increasing $\sigma$ direction.

Without loss of generality, we assume that
$\sigma$ varies from $0$ to $2\pi$, and, furthermore,
for symmetric contours we choose $\sigma$ in such a way
that $z(2\pi \! -\! \sigma , t)= \overline{z(\sigma , t)}$,
$z(0, t)=z(2\pi , t) =x_+$, $z(\pi , t) =x_-$
and $y(\sigma , t)$ is positive for $0<\sigma < \pi$.
We have: $M_+(a)=I(a)+\overline{I(\bar a)}$, where
$$
I(a)=\frac{1}{\pi i} \int_{0}^{\pi}
\frac{y(\sigma )\, dz(\sigma )}{z(\sigma )-a}\,.
$$
A straightforward calculation gives:
$$
\begin{array}{ll}
i\pi \p_t I(a) & = \, \displaystyle{
\int_{0}^{\pi}\left (
\frac{\p_t y \p_{\sigma}z +y \p_{t}\p_{\sigma}z}{z-a}
- \frac{y\p_{t}z \p_{\sigma}z}{(z-a)^2}\right )d\sigma}
\\ &\\
&=\, \displaystyle{
\int_{0}^{\pi}\left (
\frac{\p_t y \p_{\sigma}z +y \p_{t}\p_{\sigma}z}{z-a}\, d\sigma
+ y\p_{t}z \, d\left ( \frac{1}{z-a}\right )\right )}
\\ &\\
&=\, \displaystyle{
\int_{0}^{\pi}
\frac{\p_t y \p_{\sigma}z -\p_t z \p_{\sigma}y}{z-a}\, d\sigma
+ \left. \frac{y(\sigma )\p_{t}z(\sigma )}{z(\sigma )
-a}\, \right |_{0}^{\pi}}\,.
\end{array}
$$
The last term obviously vanishes and we obtain, using (\ref{appA2}):
$$
\p_t I(a) =\frac{1}{\pi i}
\int_{\gamma}\frac{v_n (z) \, |dz|}{a-z}\,.
$$
Adding $\overline{\p_t I(\bar a)}$, we get
(\ref{appA1}).

\subsection*{Appendix B}

In this Appendix we outline the proof of the following
proposition.

\begin{itemize}
\item[]
{\bf Proposition 1.}
Let $\DD$ be a compact domain bounded by a
closed piecewise analytic contour $\Gamma =\p \DD$ in the plane
with a finite number of corner points.
Consider the function $h(z)$ defined by the Cauchy-type integral
\beq\label{C1}
h(z)=\frac{1}{2\pi i}\oint_{\Gamma}
\frac{\rho (\xi )\, |d \xi |}{z-\xi}\,,
\eeq
where $\rho (\xi )$ is a bounded real-valued piecewise continuous
function on $\Gamma$ such that
\beq\label{C2}
\oint_{\Gamma}\rho (\xi )\, |d\xi |=0
\eeq
and assume that $h(z)=0$ for all
$z\in \DD$. Then $\rho \equiv 0$.
\end{itemize}

One can try to prove this statement by means of the
following elementary argument.
Let $\tau (\xi )=d\xi /|d\xi |$ be the
unit tangential vector to the curve $\Gamma$ at the
point $\xi$ represented as a complex number.
If $h(z)=0$ for all
$z\in \DD$, then the properties
of Cauchy-type integrals imply that
$\rho (\xi )/\tau (\xi )$ is the boundary value
of a holomorphic function $h(z)$ in $\CC \setminus \DD$
vanishing at infinity. In fact $h(z)$ is
given by the same integral (\ref{C1}), where
$z\in \CC \setminus \DD$. Condition (\ref{C2})
tells us that the zero at infinity is of at least
second order.
Let $w(z)$ be the conformal map from $\CC \setminus \DD$
onto the unit disk such that $w(\infty )=0$
and $\displaystyle{r=\lim_{z\to \infty} zw(z)}$ is real positive. By the
well known property of conformal maps we have
$$
\frac{dz}{|dz|}\, e^{i\,
\mbox{{\footnotesize arg}} w'(z)}=\frac{dw}{|dw|}
$$
along the curve $\Gamma$. Therefore,
\beq\label{C3}
\tau (z)=i|w'(z)|\, \frac{w(z)}{w'(z)}\,,
\eeq
and we thus see that
$$
\frac{\rho (z)\, w'(z)}{i |w'(z)|w(z)}
$$
is the boundary value of the holomorphic function $h(z)$.
Since $w'(z)\neq 0$ in $\CC \setminus \DD$, the function
$$
g(z)=h(z)\, \frac{w(z)}{w'(z)}
$$
is holomorphic there with the
{\it purely imaginary} boundary value
$$
\frac{\rho (z)}{i |w'(z)|}\,.
$$
Then the real part of this function is harmonic
and bounded in $\CC \setminus \DD$ and is equal to $0$ on the
boundary. By uniqueness of a solution to the Dirichlet
boundary value problem, ${\cal R}e \, g(z)$ must be equal to $0$
identically. Therefore, $g(z)$ takes purely imaginary values
everywhere in $\CC \setminus \DD$ and so is a constant.
By virtute of
condition (\ref{C2}) this constant must be $0$ which
means that $\rho \equiv 0$.

However, this argument is directly applicable only
for purely analytic contours for which all singularities
and zeros of the function $w'(z)$ lie strictly inside it.
For contours with corners, the corner points are singular
points of the conformal map $w(z)$.
Some more work is required to make the above argument rigorous.
Below we present another proof of Proposition 1,
which makes use of some non-trivial facts about
boundary values of analytic functions and actually
works in a much more general setting than just
a finite number of corner points\footnote{I thank D.Khavinson
who suggested this proof and explained it to me.}. It takes
advantage of translating the proposition to a statement
about analytic functions in the unit disk.

\paragraph{Sketch of proof of Proposition 1.}
If $h(z)$ defined by (\ref{C1}) is identically $0$ in
$\DD$, then it is analytic in $\CC \setminus \DD$, is
$O(1/z^2)$ near infinity (because of (\ref{C2})) and
has the boundary value $\rho (z)/\tau (z)$ almost
everywhere on $\Gamma$. (In our situation ``almost everywhere"
means everywhere except a finite number of points.)
It is known \cite[chapter III]{Privalov} that $h(z)$ belongs to the
Smirnov class $E^{1}(\CC \setminus \DD )$.

Let
$z=\varphi (w)$ be the conformal map from the unit disk
onto $\CC \setminus \DD$ such that $\varphi (0)=\infty$ and
$\varphi (w)=r/w + O(1)$ as $w\to 0$ with real $r$.
(The function $\varphi (w)$ is inverse to $w(z)$ introduced
above.) Set $\tilde h(w)=h(\varphi (w))$.
Since
$$
\tau (\varphi (w))=\frac{dz}{|dz|}
=\frac{d\varphi (w)}{|d\varphi (w)|}=
\frac{\varphi '(w)dw}{|\varphi '(w)||dw|}=iw\,
\frac{\varphi '(w)}{|\varphi '(w)|}
$$
it follows from
the above that $\tilde h(w)$ is analytic
in the unit disk with the boundary value
\beq\label{C5}
\tilde h(w)=\frac{\rho (\varphi (w))|\varphi '(w)|}{iw\, \varphi '(w)}
\eeq
almost everywhere on the unit circle
and has zero of at least
second order at $w=0$. Clearly, the function
$\tilde H(w)=\tilde h(w)\varphi '(w)$ is analytic in the unit disk
because the second order pole of $\varphi '(w)$ at $w=0$
is canceled by the zero of $\tilde h(w)$.
Furthermore, according to the Keldysh-Lavrentiev theorem
\cite[chapter 10]{Duren}, $\tilde H(w)$ belongs to the
Hardy class $H^1$. But then
the function $H(w)=w\tilde H(w)$ belongs to the same Hardy class and
takes {\it purely imaginary boundary values}
$$
-i\rho (\varphi (w))|\varphi '(w)|
$$
almost everywhere on the unit circle. The characteristic
property of functions from the class $H^1$ is that they can
be represented by the Poisson integral of their
boundary values with real positive Poisson
kernel (see \cite[chapter II, $\S 5$]{Privalov}
or Theorem 3.9 in \cite{Duren}).
This means that $H(w)$ must be
purely imaginary everywhere inside the unit disk and hence
must be a constant. Since $H(0)=0$, the constant is $0$,
so $h(z)$ vanishes identically and $\rho \equiv 0$.

The proof extends word for word to a more general case when
$\rho$ is an integrable function with respect to $|dz|$ (not
necessarily bounded) and the boundary of $\DD$ is a rectifiable
Jordan curve. It is crucial that the differential $\rho (z)|dz|$
is real valued on the boundary. If one dropped that assumption,
the statement is false. Moreover, functions from the class $E^1$
can have real boundary values in domains with cusps (see an
example in \cite{Kh}).

\section*{Acknowledgments}

The author thanks D.Kha\-vin\-son,
I.Kri\-che\-ver, M.Mi\-ne\-ev-\-Wein\-stein,
T.Ta\-ke\-be, D.Va\-si\-li\-ev and P.Wi\-eg\-mann for discussions
and D.Kha\-vin\-son for reading the manuscript.
He is also grateful to organizers of the
workshops ``Laplacian growth and related topics" (CRM, Montreal, August 2008)
and ``Geometry and integrability in mathematical physics"
(CIRM, Luminy, September 2008),
where these results were reported.
This work was supported in part
by RFBR grant 08-02-00287,
by grant for support of scientific schools
NSh-3035.2008.2 and by the ANR project GIMP No. ANR-05-BLAN-0029-01.

\end{document}